\documentclass[lnbip]{svmultln}
\usepackage{array}

\usepackage{mdwmath}
\usepackage{mdwtab}

\usepackage{flushend}           
\usepackage{fixltx2e}

\usepackage{url}
\usepackage{times}
\usepackage[pdftex]{graphicx}	
\usepackage[numbers,sort]{natbib}
\usepackage{hyperref}

\usepackage{versions}

\usepackage[capitalize]{cleveref}

\usepackage[caption=false,font=footnotesize]{subfig}

\usepackage{paralist}           

\usepackage{pifont}             

\usepackage{algorithmic}       
\usepackage{listings}           
\usepackage{moreverb}           

\usepackage{fancyvrb}           

\usepackage{eurosym}
\usepackage{units}              

\usepackage{booktabs}
\usepackage{multicol}           
\usepackage{longtable}          
\usepackage{ctable}             



\usepackage{rotating}           

\usepackage{multirow}

\newcommand{\PreserveBackslash}[1]{\let\temp=\\#1\let\\=\temp}

\usepackage[normalem]{ulem}     

\usepackage{fp}                 
\usepackage{xspace}

\newcommand{\TeamA}{TeamA\xspace}
\newcommand{\Ocuco}{PracMed\xspace}
\newcommand{\AgileTeam}{Scrum Team\xspace}
\newcommand{\TeamMember}{Scrum Team Member\xspace}
\newcommand{\TeamMembers}{Scrum Team members\xspace}
\newcommand{\ProductOwner}{Product Owner\xspace}
\newcommand{\ProductOwners}{Product Owners\xspace}

\newcommand{\ScrumMaster}{Scrum Master\xspace}
\newcommand{\ScrumMasters}{Scrum Masters\xspace}

\newcommand{\ProjectManager}{Project Manager\xspace}
\newcommand{\ProjectManagers}{Project Managers\xspace}

\newcommand{\ProductManager}{Product Manager\xspace}

\newcommand{\NumActivities}{ten\xspace}
\newcommand{\NumRoles}{eight\xspace}

\begin{document}

\mainmatter              


\subtitle{\small{Please cite as: \textcolor{red}{Noll, J., Razzak, M. A., Bass, J. M., \& Beecham, S. (2017, November). A study of the Scrum Master’s role. In International Conference on Product-Focused Software Process Improvement (pp. 307-323). Springer, Cham.}}}
\title{A Study of the \ScrumMaster's Role}

\titlerunning{Scrum Master}  
%
\author{John Noll\inst{1} \and Mohammad Abdur Razzak\inst{2}  \and Julian M Bass\inst{3} \and Sarah Beecham\inst{2} }
\authorrunning{John Noll et al.}   
%
\tocauthor{John Noll, Mohammad Abdur Razzak, Julian M. Bass, Sarah Beecham}
\institute{
  University of East London, University Way, London, E16 2RD, UK \\ 
  \email{j.noll@uel.ac.uk}
  \and
  Lero, the Irish Software Research Centre, University of Limerick, Ireland\\
  \email{{abdur.razzak, sarah.beecham}@lero.ie} 
  \and
  University of Salford, 
  The Crescent, Salford, M5 4WT, UK \\
  \email{j.bass@salford.ac.uk} 
}

\maketitle              

\begin{abstract}
Scrum is an increasingly common approach to software development adopted by organizations around the world. However, as organizations transition from traditional plan-driven development to agile development  with Scrum, the question arises as to which
Scrum role (\ProductOwner, \ScrumMaster, or \TeamMember) corresponds to a \ProjectManager, or conversely which Scrum role
should the \ProjectManagers adopt?

In an attempt to answer this question, we adopted a mixed-method research approach comprising a systematic literature review and a case study of a commercial software development team. Our research has identified activities that comprise the \ScrumMaster role, and which additional roles are actually performed by \ScrumMasters in practice.

We found \NumActivities activities that are performed by
\ScrumMasters. In addition, we found that \ScrumMasters also perform other roles, most importantly as \ProjectManagers.  This latter situation results in tension and conflict of interest that could have a negative impact on the performance of the team as a whole.


These results point to the need to re-assess the role of
\ProjectManagers in organizations that adopt Scrum as a development
approach.  We hypothesize that it might be better for \ProjectManagers
to become \ProductOwners, as aspects of this latter role are more consistent with the traditional responsibilities of a \ProjectManager.

\end{abstract}

\section{Introduction}\label{sec:intro}

Scrum~\cite{schwaber1995scrum, schwaber2002agile} is an
increasingly common approach to software development adopted by
organizations around the world.  According to the annual State of
Agile Survey~\cite{state-of-agile-survey}, \emph{94\%}
of organizations surveyed practice agile development.


However, while the vast majority of \emph{organizations} are moving
towards form of agile development, for most of these organizations,
more than half of their teams are still following traditional, plan-driven
methods~\cite{state-of-agile-survey}.  Therefore, as organizations
transition from traditional plan-driven development to agile
development with Scrum, the question arises as to which Scrum role
(\ProductOwner, \ScrumMaster, or \TeamMember) is the \ProjectManager,
or conversely which Scrum role should \ProjectManagers adopt?


In an attempt to answer this question, we used a mixed method research
approach comprising a systematic literature review, and a case study
of a commercial software development organization. Firstly, we
reviewed the literature on agile software development in order to
identify which activities are conventionally performed by
\ScrumMasters.  Then, we conducted observations and practitioner
interviews in order find out which activities are actually performed,
and which additional roles \ScrumMasters perform.


We found \NumActivities activities that are performed by
\ScrumMasters.  Of these, only three are conventional \ScrumMaster
activities.  Others would traditionally be considered the
responsibility of the \ProductOwner or \AgileTeam.  In addition, we
found that \ScrumMasters also double in other roles, most importantly
as \ProjectManagers.  This latter situation results in tension and
conflict of interest that could have a negative impact on the
performance of the team as a whole.


These results point to the need to re-assess the role of
\ProjectManagers in organizations that adopt Scrum as a development
approach.  We suggest that it might be better for \ProjectManagers
to become \ProductOwners, as this latter role is more consistent with
the traditional responsibilities of a \ProjectManager.


The rest of this paper is organized as follows: in the next section,
we present the background related to Scrum and  Scrum roles.  Next, we
describe our research method.  Following that, in \cref{sec:results}
we present our results, and a discussion of those results in
\cref{sec:discussion}.  \cref{sec:conclusions} ends with our conclusions.

\section{Background}\label{sec:background}

There are three key roles defined in the Scrum development approach:
the self-organizing \AgileTeam of developers, the \ScrumMaster, and the \ProductOwner \cite{schwaber2002agile}.
The \ProductOwner represents the external stakeholder interests (customer, users, product management) and so is the primary interface between these stakeholders and the software development team \cite{schwaber2004agile}.
The \AgileTeam is responsible for the actual software development.
A further role, Product Manager, who ``defines initial content and timing of the
release, then manages their evolution as the project progresses and variables change\ldots [and] deals with backlog, risk, and release content'' was also described in the original description of Scrum~\cite{schwaber1995scrum}; 
this role is mostly performed by the \ProductOwner in modern versions of Scrum~\cite{Cohn_2003_Need}.

The \ScrumMaster is responsible for facilitating the development
process, ensuring that the team uses the full range
of appropriate agile values, practices and rules. The \ScrumMaster 
conducts daily coordination meetings and removes any
impediments that the team encounters \cite{schwaber2002agile}. 
Six \ScrumMaster activities have been identified in a large-scale distributed organisational context: process anchor, stand-up facilitator, impediment
remover, sprint planner, scrum of scrums facilitator
and integration anchor \cite{bass2014scrum}. 
The process anchor nurtures adherence
to agile methods. The stand-up facilitator ensures that team
members share status and impediment information during each
sprint. The impediment remover ensures developers can make
progress with their work. The sprint planner supports the
user story triage and workload planning that occurs prior to
development work starting in each sprint. The scrum of scrums
facilitator coordinates work with the other \ScrumMasters in the
development program. The integration anchor facilitates the
merging of code bases developed by cooperating teams working
in parallel.

According to Schwaber and Sutherland's Scrum guidelines, ``the Scrum Master is a servant-leader for the Scrum Team. The Scrum Master helps those
outside the Scrum Team understand which of their interactions with the Scrum Team are helpful
and which aren't. The Scrum Master helps everyone change these interactions to maximize the
value created by the Scrum Team'' \cite{schwaber2016scrum}; in summary, the \ScrumMaster serves the development team. 
This is in contrast to the \ProductOwner, who is responsible for maximizing the value of the product and the work of the \AgileTeam. 
Schwaber and Sutherland \cite{schwaber2016scrum} state that although there is great flexibility in how this is achieved, the \ProductOwner is the sole person responsible for managing the Product Backlog. 

According to Schwaber and Sutherland \cite{schwaber2016scrum}, Product Backlog management tasks include: 
\begin{inparaenum}
``\item ordering the items in the Product Backlog to best achieve goals and missions;
\item optimizing the value of the work the Development Team performs;
\item ensuring that the Product Backlog is visible, transparent, and clear to all, and shows what the Scrum Team will work on next; and,
\item ensuring the Development Team understands items in the Product Backlog to the level needed.~\cite{schwaber2016scrum}''
\end{inparaenum}

Evidence from practice shows that the \ScrumMaster role is evolving.  
For example, the role is sometimes shared, and activities performed by the \ScrumMaster are varied and somewhat different from the original vision.
This was observed by Gupta et al \cite{gupta2016adapting}, who found that the challenges of adapting Scrum in a globally distributed team were helped by more than one person sharing the \ScrumMaster and \ProductOwner roles.  
Gupta et al developed a new \ScrumMaster taxonomy in which three new roles were created to reflect the complexity involved in managing a global software development team, and transitioning from Waterfall to Scrum, the roles were: \ScrumMaster cum Part \ProductOwner (where development leads were also acting in part as product owners), Bi-Scrum Master (where a development leads worked remotely with the development team) and Chief \ScrumMaster (fulfilling the need to co-ordinate among scrum teams).

According to the ISO/IEC/IEEE standard on user documentation in agile \cite{iso2012systems} the \ScrumMaster and \ProductManager have similar responsibilities when it comes to explaining  changing or new requirements. 
``The scrum master and information development lead or project manager should provide guidance to the
technical writers and other members of the agile development teams on how to handle changing or new
requirements.''  
Perhaps this conflating of roles is largely due to organizations converting the traditional project manager role to a \ScrumMaster role,  ``As more and
more of our Project Managers become Scrum Masters
and the Portfolio Managers becomes the Group Scrum
Master, our Portfolio Management Office needed to
become Agile itself \cite{tengshe2007establishing}.''

Adapting Scrum roles and creating new roles to manage large scale projects is observed in other studies, where an `Area Product Owner' (APO) role was created; this APO role was shared by two people: a system architect and a product management representative.  
The system architect worked closely with the team, while the product management representative did not interact directly with the teams \cite{paasivaara2011scaling}. This combined role (shared between two people) worked well for this organisation and was reported as one of the successes of the project. 
However, in a later study, the same authors noted that line managers had a double role: that of Scrum Master, and that of traditional line management duties involving personnel issues such as performance evaluation.
Over use of the \ScrumMaster role, who acted as a team representative at common meetings rather than rotate the role, was found problematic. The team felt that these meetings were a waste of time, and sent the \ScrumMaster instead of taking turns \cite{paasivaara2016scaling}. The frequent meetings in Scrum were also a problem in \cite{stray2013obstacles}. A \ScrumMaster's role is to facilitate daily coordination meetings where coordination meetings are used to communicate status of
development work within the team and to product owners.  
However, the efficacy of daily coordination meetings was often compromised by too many stakeholders attending, or because the meetings
were held too frequently to be beneficial for attendees \cite{stray2013obstacles}. 

Corrupting the careful balance between Scrum roles leads to other problems.  
For example Moe et al \cite{moe2008understanding} observed that the \ScrumMaster also did estimates and did not involve all the team in discussing a task. 
This lead to developers working alone, poor team cohesion, and problems emerging at the end of the sprint rather than at the beginning. 
A lack of thorough discussion was said to reduce the validity of the common backlog ``making the
developers focus more on their own plan. Since the
planning had weaknesses and none of the developers
felt they had the total overview, this probably was one
of the reasons for design-problems discovered later.''

Yet, in a recent survey that looked into whether project managers still exist in agile development teams, Shashtri and Hoda were surprised to learn that 67\% of organisations surveyed reported that they still had the \ProjectManager role.
These authors call for more research into why the \ProjectManager
continues to be present on agile software development projects, and how their role may
have changed~\cite{shastri2016does}. Conventional wisdom suggests that \ProjectManagers use a
command and control style of management, whereas \ScrumMasters focus on leading and coaching \cite{berczuk2010we}. 
As such, Scrum masters
are not line managers for their sprint team members. Further,
\ScrumMasters do not assign work items to the members of
their team, since the teams are self-organising \cite{bass2014scrum}.

In summary, there is an emerging theme in the literature, namely that the original balance of \ScrumMaster, product owner and team roles are being adapted, conflated, and possibly corrupted, to suit the needs of organizations transitioning from waterfall to Scrum, or scaling Scrum to large scale organisations. The extent to which the scrum master role has changed is unknown.  Therefore, in this study we now look to the wider literature, and specifically ask two questions:

\textbf{RQ1:} What activities do \ScrumMasters perform according to the empirical literature?

\textbf{RQ2:} What other roles do \ScrumMasters perform in practice?

We ask these questions in order to establish a broader understanding of a key Scrum role that has clearly evolved since its inception in 1995 \cite{sutherland1995business} and later refinement \cite{schwaber2016scrum}, and consider whether adapting the theory proposed by Schwaber, Sutherland and  Beedle is something to be embraced or resisted.

\section{Method}\label{sec:method}

In order to address our research questions, we adopted a mixed method approach comprising a systematic literature review and a case study of a commercial software development team \cite{creswell_research_2013}. We performed a systematic literature review \cite{keele2007guidelines} to identify the set of activities and additional roles performed by \ScrumMasters. 
Then, using observations and transcripts of semi-structured interviews we undertook as part of an empirical study, we attempted to identify benefits or issues related to these activities and roles.

\subsection{Systematic Literature Review}

Our review of the literature was conducted in five steps, by two of  the authors.

\begin{figure}
\centering{
\includegraphics[width=0.95\columnwidth]{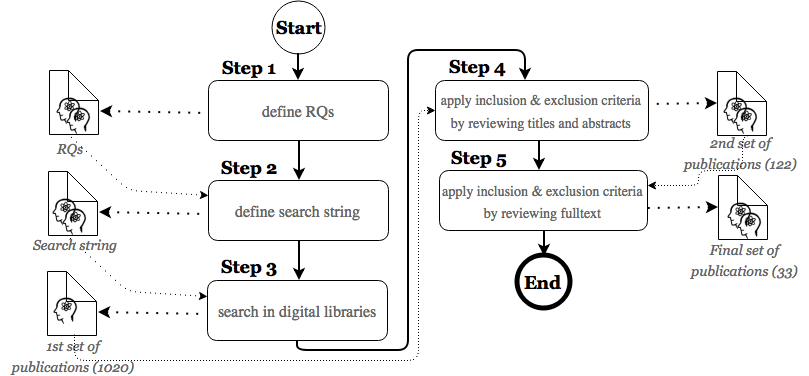}}
\caption{Systematic literature review process}\label{fig:slr-process}
\end{figure}

Two researchers were involved in the systematic literature review process (see Figure \cref{fig:slr-process}), which comprised five steps.

First, we defined two research questions:
  
\begin{enumerate}
\item What are the activities a \ScrumMaster performs?
\item What roles does the \ScrumMaster perform in addition to the \ScrumMaster role?
\end{enumerate}

Next, we defined a search string.
For expediency, we used one search string (or variants of the search string to fit the various databases) that combined both our research questions, as follows:
\begin{verbatim}
(activit* OR task* OR responsibilit* OR action* OR role* OR job*) 
AND ("Scrum Master")
\end{verbatim}

We then used this search string to search five well-established digital libraries listed in  \cref{tab:list-of-databases} for potentially relevant publications. 
This search yielded 1,020 candidate publications.
		
\begin{table}
	\caption{List of databases and number of publications.}
	\label{tab:list-of-databases}
	\centering
	\begin{tabular}{lc}
		\toprule
		\textbf{Database} & \textbf{\# of publications} \\
		\midrule
		IEEEXplore & 13\\
		ACM Digital library  & 378 \\
		Scopus & 30 \\
		Elesevier Science Direct & 282 \\
		SpringerLink & 317 \\
		\midrule
		\textbf{Total} & \textbf{1020} \\
		\bottomrule
	\end{tabular}
\end{table}

Subsequently, we applied inclusion and exclusion criteria (see \cref{tab:inclu-exclu-criteria}) to the titles and abstracts of the initial set of 1,020 publications; 
this yielded a refined set of 122 publications.  The first researcher applied the criteria, and the second researcher validated this application by independently applying the same criteria to a sample of publications.  

Finally, we again applied the inclusion and exclusion criteria to the full-text of these 122 publications, resulting in a final set of 33 publications.  In this step, both reasearchers applied the criteria independently to all 122 publications; disagreements were resolved by discussion.

\begin{table}
\caption{Inclusion and Exclusion criteria.}\label{tab:inclu-exclu-criteria}
\centering
\begin{tabular}{p{.47\textwidth}p{.5\textwidth}}
\toprule
 \textbf{Inclusion criteria} & \textbf{Exclusion criteria}  \\
\midrule
\begin{tabular}{lp{.4\textwidth}}
\textbf{IC1:} & Publication year: 2006-2017  \\
\textbf{IC2:} & Language: English \\ 
\textbf{IC3:} & Full text available and accessible \\
\textbf{IC4:} & Focus on Scrum, in the field of software engineering \\
\textbf{IC5:} & Peer reviewed work \\
\textbf{IC6:} & Answers one or more of the research questions \\
 \end{tabular} 
& 
\begin{tabular}{lp{.4\textwidth}}
\textbf{EC1:} & Is an experience report, book, presentation, or blog entry \\
\textbf{EC2:} & Is a duplicated study (where authors report similar results in two or more publications--e.g. a journal paper that is an extension of a conference paper);  exclude the least detailed paper, or if unclear include the paper that is published in the more notable venue. \\
\end{tabular}
\\
\bottomrule
\end{tabular}
\end{table}

From this final set of 33 publications, we extracted  a list of \ScrumMaster's activities and additional roles, which are reported in \cref{sec:results}.  
The first researcher extracted fragments from every paper that described \ScrumMaster activities.  
Next, the second researcher validated every one of these fragments by examining them in context to verify that each did indeed describe a \ScrumMaster activity.  
Then, working together, both researchers coalesced the validated set of fragments into \NumActivities higher level themes that represent \ScrumMaster activities.
Finally, again working together, both researchers identified

\subsection{Case Study}

The company we studied, which we will call \Ocuco, is a medium-sized Irish-based software company that develops practice and lab management software for the optical industry.

\subsubsection{Research Site}
\Ocuco employs approximately seventy staff members in its software development organization, including support and management staff.
\Ocuco's annual sales approach \euro 20 million, from customers across the British Isles, continental Europe, Scandinavia, North America and China.

Our study focused on \TeamA, who are responsible for tailoring the company's product for a large customer in North America.
The members of \TeamA are distributed over four countries on two continents, with up to eight hours difference in timezones between locations. They are using Scrum to develop their software, with two weekly sprints.
Table \cref{tab:diffteamincompanya} shows the distribution of team members; of these, two team members share the \ProductOwner role, five are developers, one is the QA/Test lead, and one is \ProjectManager.
In \TeamA, the \ProjectManager also plays role of \ScrumMaster.  
Also, the \ProductOwners report to the \ProductManager, who is based in Spain and is responsible for the strategic direction of the product.

\begin{table}[htbp]
\caption{Team Distribution.}
\label{tab:diffteamincompanya}
\centering
\begin{tabular}{lp{.5\textwidth}c}

\textbf{\emph{Country}} & {\textbf{\emph{Agile Roles}}}       & \textbf{\emph{No of Team Members}} \\
\toprule
Ireland                 & \ProductOwner                       & 1                                  \\
                        & Software Developer                  & 3                                  \\ 
                        & Quality Assurance                   & 1                                  \\
\midrule
Canada                  & \ScrumMaster (\ProjectManager)                        & 1                                  \\
                        & \ProductOwner                       & 1                                  \\ 
                        & Software Developer                  & 1                                  \\
\midrule
USA                     & Technical Lead (Software Developer) & 1                                  \\
\midrule
Spain                   & \ProductManager                     & 1                                  \\
\bottomrule
\end{tabular}

\end{table}

\subsubsection{Data Collection}

We observed \TeamA from January, 2016 through  to March, 2017.  
Specifically, one of the authors observed approximately 200 of \TeamA's Scrum ceremonies, including daily standups, sprint planning, backlog grooming, and sprint retrospectives.
Due to  team members being distributed across Europe and North America, the observations were made via video conference for each ceremony.
The same author also conducted semi-structured interviews of each member of \TeamA, which were recorded and transcribed. 
The interviews took approximately one hour, and resulted in 136 pages of transcribed verbatim data.
The interview protocol is available from \cite{Beecham_2017_Lean}.

The observer also made contemporaneous hand-written notes during both the ceremony observations and interviews. Finally, the interviewer summarized the interviews using a mind-map, and presented the result to five interviewees in an online workshop to validate the insights gained from the interviews.

\subsubsection{Data Analysis}

Interview recordings and transcripts were carefully reviewed. An open coding approach was used to identify topics in interview transcripts and contemporaneous notes of ceremonies. An approach informed by thematic analysis was used to group codes into concepts \cite{braun_using_2006}. 

\section{Findings}\label{sec:findings}\label{sec:results}
%

In this section we summarise our results and in response to our research questions, describe each of the \ScrumMaster's activities identified in our data analysis. As noted in our method, for consistency, where possible we adopt the activity name given in the literature.

\subsection{Systematic Literature Review}

Our paper selection process identified a total of 33 publications  that fit our search and inclusion criteria
(\cref{tab:pub-by-year}).

\begin{table}
  \caption{ Publication by year.}
  \label{tab:pub-by-year}
  \centering
  \begin{tabular}{lcccccccccccc}
    \toprule
    \textbf{Year} & \textbf{2006} & \textbf{2008} & \textbf{2009} & \textbf{2010} & \textbf{2011} & \textbf{2012} & \textbf{2013} & \textbf{2014} & \textbf{2015} & \textbf{2016} & \textbf{2017} & \textbf{Total} \\
    \midrule
    \textbf{Publications} & 1& 1 &1 & 3 & 4 & 4 & 1 & 8 & 2 & 6 & 2 & \textbf{33} \\
    \bottomrule
  \end{tabular}
\end{table}

\paragraph{Activities}
%
\begin{table}
\caption{\ScrumMaster activities.}\label{tab:sm-activities}
\centering
\begin{tabular}{p{.5\textwidth}ll}
\toprule
 \textbf{Activities} & \textbf{Ideal Scrum role} & \textbf{Source} \\
\toprule
Process facilitation         & \ScrumMaster  & \cite{andriyani2017reflection, baumgart2015personality, bass2014scrum, costa2014delivering} \\
Ceremony facilitation (incl. Scrum of Scrums)        &
                                                           \ScrumMaster  & \cite{bass2014scrum, baumgart2015personality,dorairaj2012understanding,alzoubi2016empirical, maranzato2011moving} \\
Impediment removal           & \ScrumMaster  & \cite{bass2014scrum, baumgart2015personality, bless2010distributed} \\
\midrule
Prioritization               & \ProductOwner & \cite{cajander2013existing, stray2011challenges} \\
\midrule
Sprint planning              & \AgileTeam    & \cite{bass2014scrum,drury2012obstacles,heikkila2015operational, vlietland2015towards} \\
Sprint reviewing             & \AgileTeam    & \cite{chamberlain2006towards, stray2016exploring} \\
\midrule
Estimation                   & \TeamMember   & \cite{daneva2013agile} \\
Integration                  & \TeamMember   & \cite{alaa2014multi,bass2014scrum} \\
\midrule
Travelling                   & none          & \cite{alzoubi2016empirical, bless2010distributed} \\
Project management           & none          & \cite{cajander2013existing, baskerville2011post, caballero2011introducing, costa2014delivering, santos2011agile} \\
\bottomrule
\end{tabular}
\end{table}

From these papers, we identified \NumActivities activities performed by
\ScrumMasters; these are shown in \cref{tab:sm-activities}.   These activities are defined as follows:

\begin{description}
\item [Process facilitation] involves guiding the \AgileTeam on how to
  use Scrum to achieve their objectives.
\item [Ceremony facilitation] involves moderation of the daily standup,
  backlog grooming, sprint planning, and   sprint retrospective
  meetings that occur during each sprint.
\item [Impediment removal] is part of the \ScrumMaster as ``servant
  manager'' role: the \ScrumMaster serves as a buffer between the
  \AgileTeam and external pressures, and also attempts to secure
  resources or remove blockers to progress that come from outside the team.
\item [Prioritization] involves ordering stories on the product and sprint
  backlogs by order of importance.
\item [Sprint planning] identifies those stories on the product
  backlog that will fit into a single sprint, taking into account team velocity and
  capacity, and story estimates.
\item [Sprint reviewing] is part of the Sprint Retrospective ceremony
  where the team identifies what went well, what could be improved,
  and might be added or removed from their process to be more effective.
\item [Estimation] assigns a value in ``story points'' or ideal
  engineering  time representing the effort required to complete a
  story.
\item [Integration] facilitates amalgamation of software elements.
\item [Travelling]  is an activity associated with distributed teams
  that involves visiting different sites where teams are located, to
  facilitate communications~\cite{bass2015how}.
\item [Project management] is a traditional management activity found
  in waterfall-style development projects.
\end{description}

\paragraph{Roles}

Fifteen papers mentioned other roles that \ScrumMasters hold in addition to that of \ScrumMaster.
These are summarized in \cref{tab:additional-roles}. 

\begin{table}
  \caption{ \ScrumMaster additional roles.}  \label{tab:additional-roles}
  \centering
  \begin{tabular}{lcc}
    \toprule
    \textbf{Role}                            & \textbf{Company-size} & \textbf{Source}                                                                                                   \\
    \midrule
    Project Manager                          & Large-scale           & \cite{gren2017group, hoda2016multi, tuomikoski2009absorbing, stray2016daily, moe2008scrum} \\
    Product Owner                            & unclear               & \cite{cajander2013existing, tuomikoski2009absorbing} \\
    Architect/Software Designer              & Large-scale           & \cite{diaz2014agile, sekitoleko2014technical} \\
    Project Lead                             & Large-scale           & \cite{diebold2015practitioners} \\
    Developer/Senior Engineer                & Large-scale           & \cite{diebold2015practitioners, garbajosa2014communication, stray2011challenges, hoda2016multi, li2010transition} \\
    Team Leader                              & Large-scale           & \cite{diebold2015practitioners, galster2016multiple, gren2017group} \\
    Test Lead                                & unclear               & \cite{hoda2016multi} \\
    Head of Department/Dir. of Eng/Dev. Mgr. & Large-scale           & \cite{alahyari2017study, stray2011challenges, hoda2016multi, vlaanderen2012growing} \\
    \bottomrule
  \end{tabular}
\end{table}

Of these \NumRoles roles, four (Architect/Software Designer, Developer/Senior Engineer, Team Leader, and Test Lead) would be considered technical roles, and three (Project Manager, Project Lead, Head of Department) are management roles.  
In total, nine of fifteen papers reported the \ScrumMaster also taking
on some kind of management role, with six explicitly mentioning ``Project Manager'' or ``Project Lead.''

\subsection{Case Study}

We observed this tension and conflict of interest in our case study organization.
On the one hand, the \ScrumMaster performs project management duties:

\begin{quote}
	\emph{
		So, we do all the traditional project management roles as in doing the scope statement, the planning, change control process, communication management plan and all that stuff. 
		And, then internally [we act as] Scrum Master.}
\end{quote}

The planning part of this role has a waterfall characteristic:

\begin{quote} 
	\emph{
		When I got to start working on this project when there was a contract -- there is a very specific set of requirements. 
		\ldots there is a very specific budget for example, and the timeline is normally set as well [at] a high level. 
	}
\end{quote}

In \Ocuco, project management for projects involving customization for large accounts, also involves interfacing with the customer.
The \ScrumMaster admits balancing these two roles creates tension:
\begin{quote} 
	\emph{\ldots{}Madness! It's hard. \ldots if you know about one role not the other, I think it's easier because you do the best you can in your \ScrumMaster role or you do the best in your \ProjectManager role ignoring the other. 
		Now, the dilemma is as a \ProjectManager I still know what the \ScrumMaster role is, I know the Agile team -- I know I am not supposed to break their rules and let them be self-organizing and do all of that. 
		At the end you have the client to answer [to], you have management to answer to. 
		So, you can't just say oh well it's in the sprint or they plan for it or I don't know when its gonna get done because team is self-organizing. 
		}
\end{quote}

In particular, there is tension between the \ProjectManager as customer interface, and the \ScrumMaster role:
\begin{quote} 
	\emph{\ldots{}Yah, pressure will always  be there  \ldots 
		An example would be, the client would want to know \ldots exactly when all [features] are going to get done. 
		Now, in an Agile world there is no way that I could tell them when they are going to get done until the estimates are there\ldots with a client it's hard because I cannot just tell them we are doing Agile. 
}
\end{quote}

\section{Discussion}\label{sec:discussion}
%

Scrum defines only three roles: \ProductOwner, \TeamMember, and
\ScrumMaster~\cite{schwaber2002agile}.  This results in a balance
between the customer, user, and other stakeholder interests, which are
represented by the \ProductOwner, and the technical realities of
software development, which are represented by the \AgileTeam.  The
\ScrumMaster facilitates the interaction between these two interests,
and also serves to insulate the team as a whole from external
distractions (hence the description ``servant-leader'' that is often
used to describe \ScrumMasters~\cite{Cohn_2003_Need}).

Three \ScrumMaster activities (Process facilitation, Ceremony facilitation, and Impediment removal) that formed part of our \NumActivities activities observed from the literature would be considered ``traditional'' \ScrumMaster activities, as defined by Schwaber and Beedle~\cite{schwaber2002agile}.
Prioritizing, on the other hand, is supposed to be the responsibility of the \ProductOwner, and Estimation is supposed to be performed by the \TeamMembers~\cite{schwaber2002agile}.
While the \ScrumMaster may \emph{facilitate} these activities, he or she is not supposed to perform them; this is because Scrum relies on a balance of power between ``business'' and ``technical'' interests in order to set realistic sprint goals~\cite{schwaber2002agile,leffingwell2007scaling}. Given the \ScrumMaster's role as facilitator, and mediator between
the \ProductOwner and the
\AgileTeam, overloading the \ScrumMaster role with project management
introduces a conflict of interest that can compromise the
\ScrumMaster's ability to ensure a balance between the interests of
external stakeholders and the \AgileTeam: the \ScrumMaster is supposed
to insulate the team and remove impediments, but as \ProjectManager,
he or she would also have responsibilities to achieve objectives set by
higher levels of the organization.  
Stray and colleagues observed that when the \ScrumMaster is viewed as a manager rather than facilitator, the daily standup becomes a management reporting exercise rather than a team communication meeting \cite{stray2013obstacles}.

\subsection{The Way Ahead}

If tensions are created when the \ScrumMaster activities are combined with \ProjectManager activities, which Scrum role is the right role to perform \ProjectManager activities?

To answer this question, it's useful to consider what project management involves in Scrum, especially considering Scrum teams are supposed to be ``self organizing.''
Schwaber defines five project management activities that must be carried out when undertaking development using the  Scrum approach:
\begin{compactenum}
\item Vision management -- establishing, nurturing, and communicating the product vision.
\item ROI management -- monitoring the project's progress against Return on Investment goals, including updating and prioritizing the product backlog to reflect these goals.
\item Development iteration management -- expanding items on the Product Backlog into items for the Sprint Backlog, then implementing those items in order of priority.
\item Process management -- facilitating ceremonies, removing impediments, and shielding the team from outside interference.
\item Release management -- deciding when to create an official release, in response to market pressures and other investment realities.
\end{compactenum}
Of these, only \emph{Process management} is the responsibility of the \ScrumMaster; \emph{Development iteration management} is the responsibility of the development team, and the remaining activities (Vision management, ROI management, and Release management) are the \emph{\ProductOwner's} responsibility. 

This suggests that, when organizations decide to adopt Scrum, their existing \ProjectManager's should be assigned to  the \ProductOwner role.
The advantages are twofold: first, as \ProductOwners, 
\ProjectManagers could advocate for business requirements without feeling tension with their \ProductOwner responsibilities, since such advocacy is consistent with the \ProductOwner role.  

Second, the \ScrumMaster would be free to support the \AgileTeam when business requirements conflict with technical reality, 
and to support the \ProductOwner when business priorities differ from \TeamMember preferences (for example, when certain mundane functionality must be developed to keep the product roadmap progressing, at the expense of more technically interesting features), 
and to support both when upper management pressure threatens to override or compromise the team's own decisions.

\subsubsection{Limitations}

Practitioner roles, such as that of \ScrumMaster, are rapidly evolving
and hence, while literature is important, it cannot be solely
relied upon for an up-to-date perspective.  On the other hand, an
empirical case study, while providing more up-to-date insights,
necessarily derives those insights from at most a handful of settings.

This research adopts a mixed method approach to compensate for the
weaknesses of each research approach used in isolation, by combining a
systematic literature review with an empirical case study in
a mixed method approach to provide a broad perspective based on the
literature that is supported by observations from a case study.

Our insights into the tensions and conflicts created by combining the \ScrumMaster and  \ProjectManager roles are based on observations of a single development team and interviews of one \ScrumMaster/\ProjectManager.  
As such, we must be extremely cautious about generalizing our results.
However, our observations do suggest two propositions that can serve as the basis for further research:

\textbf{P1:} When adopting Scrum, teams will be more successful if the former \ProjectManager adopts the \ProductOwner role rather than the \ScrumMaster role.

Conversely, 

\textbf{P2:} When adopting Scrum, teams that combine the \ScrumMaster and \ProjectManager roles will experience tension resulting from the conflict of interests between these two roles.

\section{Conclusions}\label{sec:conclusions}
%
In this study, we adopted a mixed method research approach to try to
answer two research questions:
\begin{compactenum}
\item What activities do \ScrumMasters perform according to the empirical literature?
\item What other roles do \ScrumMasters perform in practice?
\end{compactenum}

We first performed a systematic literature review related to the \ScrumMaster role and then a case study to uncover empirical evidence of what activities \ScrumMaster's
actually perform, and what additional roles they take on.
This review revealed \NumActivities activities that are performed by \ScrumMasters,
and \NumRoles additional roles that \ScrumMasters also play.

Combining the findings from the literature with observations from a case study of a medium-sized development organization, we identified 
tensions and conflicts between the \ScrumMaster role and the \ProjectManager role that are often  combined in practice.  
As such, we propose that, when adopting Scrum, organizations appoint existing \ProjectManagers to the role of \ProductOwner, rather than that of \ScrumMaster.

\section{Acknowledgments}\label{sec:acknowledgments}
We thank the members of TeamA and members of the Project Management Team for their generous and thoughtful collaboration on this study, and PracMed, for allowing us to study their software development efforts. This work was supported, in part, by Science Foundation Ireland grants
10/CE/I1855 and 13/RC/2094  to Lero - the Irish Software Research Centre
(\url{www.lero.ie}).

\bibliographystyle{splncs}
\bibliography{top}

\end{document}